\pdfoutput=1
\RequirePackage{ifpdf}
\ifpdf % We are running pdfTeX in pdf mode
\documentclass[pdftex]{sigma}
\else
\documentclass{sigma}
\fi

\numberwithin{equation}{section}

\begin{document}

\allowdisplaybreaks

\renewcommand{\thefootnote}{$\star$}

\renewcommand{\PaperNumber}{002}

\FirstPageHeading

\ShortArticleName{Symmetries and Special Solutions of Reductions of the Lattice Potential KdV Equation}

\ArticleName{Symmetries and Special Solutions of Reductions\\ of the Lattice Potential KdV Equation\footnote{This paper is a~contribution to the Special Issue in honor of
Anatol Kirillov and Tetsuji Miwa.
The full collection is available at
\href{http://www.emis.de/journals/SIGMA/InfiniteAnalysis2013.html}{http://www.emis.de/journals/SIGMA/InfiniteAnalysis2013.html}}}

\Author{Christopher M.~ORMEROD}
\AuthorNameForHeading{C.M.~Ormerod}

\Address{Department of Mathematics, California Institute of Technology, \\ 1200 E California Blvd, Pasadena, CA 91125, USA}
\Email{\href{mailto:christopher.ormerod@gmail.com}{christopher.ormerod@gmail.com}}
\URLaddress{\url{http://www.math.caltech.edu/~cormerod/}}

\ArticleDates{Received September 19, 2013, in f\/inal form December 28, 2013; Published online January 03, 2014}

\Abstract{We identify a periodic reduction of the non-autonomous lattice potential Korte\-weg--de Vries equation with the additive discrete Painlev\'e equation with $E_6^{(1)}$ symmetry. We present a description of a set of symmetries of the reduced equations and their relations to the symmetries of the discrete Painlev\'e equation. Finally, we exploit the simple symmetric form of the reduced equations to f\/ind rational and hypergeometric solutions of this discrete Painlev\'e equation.}


\Classification{39A10; 37K15; 33C05}

\renewcommand{\thefootnote}{\arabic{footnote}}
\setcounter{footnote}{0}

\section{Introduction}

Finding explicit solutions to integrable partial dif\/ferential equations in terms of solutions of ordinary dif\/ferential equations, such as the Painlev\'e equations, is a topic that is of interest to many researchers \cite{AblowitzSegur, Kruskal, FlaschkaNewell, SKdVP6I}. Finding explicit solutions to the discrete analogues of integrable partial dif\/ferential equations, integrable lattice equations, in terms of known ordinary dif\/ference equations, such as the discrete Painlev\'e equations, has recently been a hot topic \cite{Hay,hay2013systematic, Hay:dSKdVreductions, Ormerod:qP6, ormerod2013twisted,OvdKQ:reductions, Gramani:Q4eE8}.

The symmetries of the Painlev\'e equations are well known to be realizations af\/f\/ine Weyl groups~\cite{NoumiqP4}. The work of Sakai provides a geometric framework for these realizations \cite{Sakai:rational}. Another approach to symmetries of discrete Painlev\'e equations are discrete Schlesinger transformations, which can be derived by the framework of connection preserving deformations \cite{Borodin:connection, Sakai:qP6, Ormerodlattice}. This article presents a description of the symmetries of periodic reductions of quad equations. A~discussion of the symmetries of reductions is a necessary step towards identifying reductions with the full parameter versions of some of the higher discrete Painlev\'e equations.

Our second aim is to present a novel method of f\/inding special solutions of reductions in terms of the lattice equations. It is well known that the Painlev\'e equations admit rational and hypergeometric solutions \cite{Clarkson:special}. It is even more surprising that the discrete Painlev\'e equations also admit rational and hypergeometric solutions \cite{Noumi:survey}, basic hypergeometric solutions \cite{qPhypergeometric} and even elliptic hypergeometric solutions \cite{Elliptichypergeomtric, rains:isomonodromy}. There are many approaches to f\/inding special solutions, such as a direct approach \cite{Sakai:qP6}, bilinear approaches \cite{KajiwaraqP3I}, orthogonal polynomial approaches \cite{OrmerodqPV, OrmerodForresterWitte, Witte:Correspondence} and geometric approaches \cite{qPhypergeometric, Murata:Riccati}. Our approach could be called a reductive approach.

To demonstrate the general symmetry structure of reductions and our approach to f\/ind special solutions, we will consider an identif\/ication of a periodic reduction of the nonautonomous lattice potential Korteweg--de Vries equation,
\begin{gather}\label{H1}
(w_{l,m} - w_{l+1,m+1})(w_{l+1,m} - w_{l,m+1}) = p_l - q_m,
\end{gather}
with the additive Painlev\'e equation with $E_6^{(1)}$ symmetry, given by
\begin{subequations}\label{dPE6}
\begin{gather}
(\tilde{z} + y)(y + z) = \dfrac{(y-a_3)(y-a_4)(y-a_5)(y-a_6)}{(y-a_1+t)(y-a_2+t)},\\
(\tilde{y} + \tilde{z})(y + \tilde{z}) = \dfrac{(\tilde{z}+a_3)(\tilde{z}+a_4)(\tilde{z}+a_5)(\tilde{z}+a_6)}{(\tilde{z} + a_7 + t)(\tilde{z} + a_8 + t)},
\end{gather}
\end{subequations}
where $\tilde{t} = t+\delta$ and $\tilde{f} = \tilde{f}(t) = f(t+\delta)$ and
\begin{gather}\label{constraint}
(a_1 + a_2 + a_7+a_8)-(a_3 + a_4 + a_5 + a_6) =\delta.
\end{gather}
This is sometimes known as the asymmetric $\mathrm{d}$-$\mathrm{P}_{\rm IV}$ \cite{ramani2001special}, the dif\/ference $\mathrm{P}_{\rm VI}$~\cite{ArinkinBorodin}, the additive discrete Painlev\'e equation with $E_6^{(1)}$ symmetry or  $\mathrm{d}$-$\mathrm{P}\big(A_2^{(1)*}\big)$ \cite{DzhamaySakaiTakenawa, murata2004new, Sakai:rational}. The Riccati solutions were found relatively early by Ramani et al.~\cite{ramani2001special} and their expressions in terms of ${}_3F_2(1)$ functions was presented by Kajiwara~\cite{kajiwara:hypergeometricEs}.

There are very good reasons to consider this equation. Firstly, this equation possesses the sixth Painlev\'e equation as a continuum limit and is the lowest member of the additive type discrete Painlev\'e equations that does not arise as a contiguous relation for a continuous Painlev\'e equation~\cite{Noumi:survey}.

Secondly, it is very interesting that this equation arises from two characteristically distinct Lax pairs; a dif\/ference-dif\/ference Lax pair, by Arinkin and Borodin \cite{ArinkinBorodin}, and a recent dif\/ferential-dif\/ference Lax pair, by Dzhamay et al.~\cite{DzhamaySakaiTakenawa}. These two Lax pairs also arise from two distinct notions of isomonodromy. The f\/irst known Lax pair was derived as a discrete analogue of an isomo\-nodromic deformation in that it preserves a connection matrix, in the sense of Birkhof\/f~\mbox{\cite{Birkhoff, Borodin:connection}}. The second was derived as a discrete isomonodoromic deformation of a $3\times 3$ system, i.e., a~Schlesinger transformation, such as those considered by Jimbo and Miwa~\cite{Jimbo:Monodromy2}.

Finally, while the $q$-Painlev\'e equation with $E_6^{(1)}$-symmetry was recently identif\/ied as a reduction of the lattice Schwarzian Korteweg--de Vries equation~\cite{OvdKQ:reductions}, the relation between~\eqref{dPE6} and any lattice equation is not known at this point. These two systems have a rich enough symmetry structure to extrapolate the general symmetry structure for more general reductions of lattice equations. This work on symmetries and special solutions is applicable to a wide class of reductions, and includes those presented in \cite{OvdKQ:reductions} and degenerations thereof.

The outline of the paper is as follows: in Section~\ref{sec:red} we will specify the reduction and its Lax representation, in Section~\ref{sec:corr} we will discuss the correspondence between the reduction and \eqref{dPE6}, in Section~\ref{sec:sym} we engage in a discussion of the symmetries of the reduction, which are described in terms of \eqref{H1}, and their relation to the symmetries of \eqref{dPE6} and in Section~\ref{sec:sols} we will discuss the way in which the reduction leads us to a fundamental set of rational and Riccati solutions.

\section{Reduction}\label{sec:red}

The discrete potential KdV equation, given by \eqref{H1}, was one of the f\/irst integrable lattice equations to be derived \cite{Nijhoff:lkdvreview,Nijhoff:dSKdV}. Autonomous reductions of this equation have been considered by many authors \cite{hone2013integrability,  Nijhoff:dSKdVP6, dKdVreds, Dinh, vdKRQ:Staircase}, and there have been some studies of non-autonomous similarity reductions \cite{Nihoff:SimilarityqP2, OvdKQ:reductions}. In~\cite{OvdKQ:reductions}, the equation was treated as a test case, where we obtained a dif\/ference version of $\mathrm{d}$-$\mathrm{P}_{\rm IV}$. Here, we will explore a much more involved reduction.

We will consider periodic $(4,2)$-reductions, which are special solutions satisfying the constraint
\begin{gather}\label{periodic}
w_{l+4,m+2} = w_{l,m}.
\end{gather}
A special case of these solutions are $(1,2)$-reductions, which may be expressed in terms of the solution of a discrete analogue of the f\/irst Painlev\'e equation and a version of the discrete fourth Painlev\'e equation \cite{OvdKQ:reductions}. In order to specify a $(4,2)$-reduction, we are required to specify six initial conditions, which are periodically continued in both directions via~\eqref{periodic}, making this a~six-dimensional mapping that we will f\/ind a suf\/f\/icient number of transformations and integrals to express as a two-dimensional mapping.

\begin{figure}[t]\centering
\includegraphics{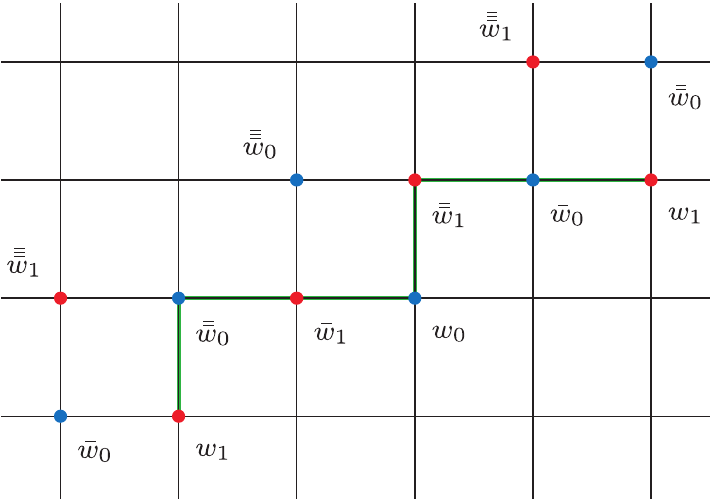}
\caption{The labeling of lattice points in accordance with the prescribed periodicity and def\/inition of $n = 2m-l$.}\label{labelling}
\end{figure}

We follow \cite{OvdKQ:reductions} by def\/ining an evolution variable, $n$, by
\begin{gather}\label{ndef}
n = 2m-l.
\end{gather}
For every value of $n$, up to periodicity, we have two distinct lattice values, $w_{0,n}$ and $w_{1,n}$. In this way, every point $(l,m) \in \mathbb{Z}^2$, is associated with either a value, $w_{0,n}$ or $w_{1,n}$. For convenience, we will use the notation $w_{i,n} = w_i$ and $w_{i,n+1} = \bar{w}_i$. We have shown this is Fig.~\ref{labelling}.

In order for the constraint to def\/ine the iterates of $w_{0}$ and $w_{1}$ consistently, we require the equation def\/ining the evolution at $(l,m)$ to coincide with the equation at $(l+4,m+2)$, hence, we require
\begin{gather*}
p_{l+4}- q_{m+2} = p_l - q_m,
\end{gather*}
which, by a separation of variables argument, def\/ines a constant in $l$ and $m$, which we label $h$, given by
\begin{gather*}%\label{varconst}
p_{l+4}- p_l = q_{m+2} - q_m := h.
\end{gather*}
This dif\/ference equation def\/ines $6$ additional constants. That is to say that the dif\/ference equation for $p_l$ is of degree $4$, def\/ining $4$ constants in general, and the dif\/ference equation for $q_m$ is of degree $2$, def\/ining $2$ new constants. We label these constants $a_1, \ldots, a_6$, which enter the system via $q_m$ and $q_l$ by letting
\begin{gather}
\label{qmdef} q_m  =   \begin{cases}
\dfrac{mh}{2}-a_1 & \text{if} \  m = 0\  \mathrm{mod}\, 2,\vspace{1mm}\\
\dfrac{mh}{2}-a_2 & \text{if} \  m = 1\  \mathrm{mod}\, 2,
\end{cases}  \\
\label{pldef} p_l  =   \begin{cases}
\dfrac{lh}{4}-a_3 & \text{if} \  l = 0 \  \mathrm{mod}\, 4,\vspace{1mm}\\
\dfrac{lh}{4}-a_4 & \text{if} \  l = 1 \  \mathrm{mod}\, 4,\vspace{1mm}\\
\dfrac{lh}{4}-a_5 & \text{if} \  l = 2 \  \mathrm{mod}\, 4,\vspace{1mm}\\
\dfrac{lh}{4}-a_6 & \text{if} \  l = 3 \  \mathrm{mod}\, 4.
\end{cases}
\end{gather}
The mapping that brings $n \to n+1$ is called the generating shift, as any shift on the lattice is some power of the generating shift. The generating shift is equivalent to the shift $(l,m) \to (l+1,m+1)$, hence, the mapping corresponding to $n \to n+1$ permutes the roles of the~$a_i$ in the following way
\begin{subequations}\label{systemw}
\begin{gather}\label{parchange}
\begin{pmatrix}
\bar{a}_1 & \bar{a}_2 & \bar{a}_3 \\
\bar{a}_4 & \bar{a}_5 & \bar{a}_6
\end{pmatrix} =
\begin{pmatrix}
a_2 +\dfrac{h}{2} & a_1 - \dfrac{h}{2} & a_4 + \dfrac{h}{4} \vspace{1mm}\\
a_5 + \dfrac{h}{4} & a_6 + \dfrac{h}{4} & a_3 - \dfrac{3h}{4}
\end{pmatrix}.
\end{gather}
In particular, notice that $\bar{\bar{\bar{\bar{a}}}}_i = a_i$.
If we assume $l$ and $m$ are $0$ $\mathrm{mod}\, 4$, under this correspondence, the equations governing the lattice variables are
\begin{gather}
\label{wevol1}(\bar{w}_0 - \bar{\bar{w}}_0)(w_1 -  \bar{\bar{\bar{w}}}_1)  = a_1-a_6 - \dfrac{nh}{4}-h,\\
\label{wevol0}(\bar{w}_1 - \bar{\bar{w}}_1)(w_0 -  \bar{\bar{\bar{w}}}_0)  = a_2-a_4 - \dfrac{nh}{4}.
\end{gather}
\end{subequations}
Let us simplify these equations by introducing variable $u$ and $v$ by
\begin{gather*}
u =  \bar{w}_0 - w_0,\qquad
v =  \bar{w}_1 - w_1,
\end{gather*}
hence, \eqref{wevol1} and \eqref{wevol0} become the degree 4 mapping
\begin{subequations}\label{systemuv}
\begin{gather}
\bar{u}(v + \bar{v} + \bar{\bar{v}}) = a_1 - a_6 - \dfrac{nh}{4} - h,\\
\bar{v}(u + \bar{u} + \bar{\bar{u}}) = a_2 - a_4 - \dfrac{nh}{4}.
\end{gather}
\end{subequations}
We use the second equation to obtain $v$ in terms of $u$ and its iterates, which then gives us an equation which may be written solely in terms of $\nu = \bar{u}/u$, reducing this fourth order system to a third order system
\begin{gather*}
\dfrac{a_1 - a_5 - \frac{n+4}{h}}{1+\bar{\nu} + \bar{\nu} \bar{\bar{\nu}}} + \dfrac{\nu \left( a_2 - a_4 - \frac{n h}{4}\right)}{1+\nu + \nu \bar{\nu}} + \dfrac{\nu \underline{\nu} \left(a_1-a_3 - \frac{nh}{4} \right)}{1 + \underline{\nu} + \nu \underline{\nu}} = a_1-a_6 - \dfrac{(n+4)h}{4}.
\end{gather*}
It is not completely trivial to see that this is  a third order system with an invariant, $d_1$, which def\/ines our second order evolution
\begin{gather}\label{eqnu}
\dfrac{\nu \underline{\nu} \left(a_1 - a_3 - \frac{nh}{4} \right)}{1+\underline{\nu} + \underline{\nu} \nu} - \dfrac{a_2 - a_4 - \frac{nh}{4}}{1+ \nu + \nu \bar{\nu}} =  d_1 + a_1 + a_4 + a_5.
\end{gather}
It is worth reminding the reader that we are primarily interested in the fourth power of this map, because the parameters change with each power of the map in accordance with \eqref{parchange}, in a similar manner to the asymmetric forms of discrete Painlev\'e equations provided by Kruskal et al.~\cite{Kruskal:AsymmetricdPs}. To avoid any confusion, we use a common notation to describe this map:{\samepage
\[
\left(
\begin{array}{@{}c c c c}
a_1 & a_2 & a_3 & d_1 \\
a_4 & a_5 & a_6 & d_2
\end{array}; n,\underline{\nu}, \nu\!
\right) \to
\left(
\begin{array} {@{}c c c c}
a_2 +\dfrac{h}{2} & a_1 - \dfrac{h}{2} & a_4 + \dfrac{h}{4} & d_1\vspace{1mm}\\
a_5 + \dfrac{h}{4} & a_6 + \dfrac{h}{4} & a_3 - \dfrac{3h}{4} & d_2
\end{array}; n+1, \nu, \overline{\nu}\!
\right).
\]
The fourth power of this mapping f\/ixes the $a_i$ variables and sends $n$ to $n+4$.}

We now describe a Lax pair for this system. A general method for determining the Lax representation of an autonomous reduction has been known for some time. While there were examples of derivations of Lax pairs for non-autonomous lattice equations~\cite{Hay:Hierarchies, Hay}, a direct method for determining a Lax representation was presented only recently~\cite{Ormerod:qP6, OvdKQ:reductions}. One of the interesting consequences of this theory is that the resulting Lax matrices factorize in a novel manner.

The starting point for the Lax pair for the reduction is the Lax pair for the lattice equation. In this case, the Lax representation is given by
%\begin{subequations}
\begin{gather*}
\Psi_{l+1,m} =  L_{l,m} \Psi_{l,m},\\
\Psi_{l,m+1} =  M_{l,m} \Psi_{l,m},
\end{gather*}
%\end{subequations}
where
\begin{subequations}
\begin{gather}
\label{LaXL}
L_{l,m} =  \begin{pmatrix}
 -w_{l+1,m} & 1 \\
 -\kappa -w_{l,m} w_{l+1,m}+p_l & w_{l,m}
\end{pmatrix}, \\
\label{LaXM}
M_{l,m} =  \begin{pmatrix}
 -w_{l,m+1} & 1 \\
 -\kappa -w_{l,m} w_{l,m+1}+q_m & w_{l,m}
\end{pmatrix},
\end{gather}
\end{subequations}
where $\kappa$ is a spectral parameter. The consistency of this system, in the calculation of $\Psi_{l+1,m+1}$, gives us the condition
\begin{gather*}%\label{CompH1}
L_{l,m+1}M_{l,m} = M_{l+1,m}L_{l,m},
\end{gather*}
which is equivalent to imposing \eqref{H1}.

We introduce a new non-autonomous spectral paramater, related to $\kappa$ by
\begin{gather}\label{xdef}
x = \dfrac{lh}{4} - \kappa.
\end{gather}
Using \eqref{ndef}, \eqref{xdef}, \eqref{qmdef} and \eqref{pldef}, we can represent linear problems in $l$, $m$ and $\kappa$ in terms of~$x$,~$n$ and the $a_i$ variables. Let us form two operators, $A_n(x)$ and $B_n(x)$, that are equivalent to the shifts~$(l,m) \to (l+4,m+2)$ and $(l+1,m+1)$ respectively. This gives us a linear system satisfying the equations
\begin{subequations}
\begin{gather}
\label{Ydiff}Y_n(x+h)  = A_n(x) Y_n(x),\\
Y_{n+1}\left(x+\dfrac{h}{4}\right)  = B_n(x) Y_n(x).
\end{gather}
\end{subequations}
These matrices are given by the products
\begin{gather*}
A_n(x) \cong  L_{l+3,m+2}L_{l+2,m+2}M_{l+2,m+1}L_{l+1,m+1}L_{l,m+1}M_{l,m},\\
B_n(x) \cong  L_{l,m+1}M_{l,m}.
\end{gather*}
Explicitly, this is
\begin{gather}
%\label{prodBw}
\nonumber
B_n(x)  =\begin{pmatrix}
 -\bar{w}_1 & 1 \\
 x-a_3-\bar{\bar{w}}_0 \bar{w}_1 & \bar{\bar{w}}_2
 \end{pmatrix}
 \begin{pmatrix}
 -\bar{\bar{w}}_0 & 1 \\
 x+\dfrac{n h}{4}-a_1 -\bar{\bar{w}}_0w_1 & w_1
 \end{pmatrix},\\
A_n(x)  = \begin{pmatrix}
 -w_1 & 1 \\
 x-a_6 -\bar{w}_0 w_1 & \bar{w}_0
 \end{pmatrix}
 \begin{pmatrix}
 -\bar{w}_0 & 1 \\
 x-a_5 -\bar{w}_0 \bar{\bar{w}}_1 & \bar{\bar{w}}_1
 \end{pmatrix} \nonumber\\
 \hphantom{A_n(x)  =}{}\times
 \begin{pmatrix}
 -\bar{\bar{w}}_1 & 1 \\
 x+\dfrac{nh}{4} - a_2 - w_0\bar{\bar{w_{1}}} & w_0
 \end{pmatrix}
 \begin{pmatrix}
 -w_0 & 1 \\
 x-a_4 - w_0 \bar{w}_1 & \bar{w}_1
 \end{pmatrix}B_n(x),
 \label{prodAw}
\end{gather}
whose compatibility, given by
\begin{gather*}
A_{n+1}\left( x+ \dfrac{h}{4} \right) B_n(x) - B_n(x+h)A_n(x) \equiv 0,
\end{gather*}
gives \eqref{systemw}. Through gauge transformations, we could express this in terms of the~$u$ and~$v$, or~$\nu$, but we leave this out for succision.

\section[Correspondence with d-P$\big(A_2^{(1)*} \big)$]{Correspondence with d-P$\boldsymbol{\big(A_2^{(1)*} \big)}$}\label{sec:corr}

For very simple reductions, it is often straightforward to write down a correspondence between the variables on the lattice and the variables of the corresponding discrete Painlev\'e equation. For higher equations, the correspondences may be highly non-trivial and requires some auxilia\-ry information as a guide. In the case of $q$-$\mathrm{P}\big(A_2^{(1)}\big)$, the auxiliary information was the Lax pair, which was found by Sakai \cite{SakaiE6} and Yamada \cite{Yamada:LaxqEs}. Nicholas Witte and this author showed these two Lax pairs were related, furthermore that the associated linear problem for the special hypergeometric solutions may be expressible in terms of a certain orthogonal polynomial ensemble~\cite{Ormerod:qE6}.

Our guide in this case is the Lax pair of Arinkin and Borodin \cite{ArinkinBorodin}. While the result of Arinkin and Borodin was indeed of interest to many in the f\/ield, the Lax pair presented was not as explicitly presented as other Lax pairs in the literature (see for example \cite{Sakai:qP6, Murata2009, Gramani:Isomonodromic}). We use this opportunity to present an explicit parameterization of this Lax pair.

Before doing so, we f\/irst examine some of the properties of the Lax pair we have thus far. The matrix $A_n(x)$ is of degree $3$, and may be written
\begin{gather*}
A_n(x) = \mathcal{A}_{0,n} + \mathcal{A}_{1,n} x + \mathcal{A}_{2,n} x^2 + \mathcal{A}_{3,n} x^3,
\end{gather*}
where $\mathcal{A}_{3,n} = I$. The other property that is important is that, by taking the determinant of~\eqref{prodAw}, we see that
\begin{gather}\label{detrel}
\det A_n(x) = \left(x-a_1+\dfrac{nh}{4}\right)\left(x-a_2+\dfrac{nh}{4}\right)(x-a_3)(x-a_4)(x-a_5)(x-a_6).
\end{gather}
What is crucial to making the correspondence is $\mathcal{A}_{2,n}$, which is of the form
\[
\mathcal{A}_{2,n} = \begin{pmatrix}
d_1 + \dfrac{nh}{4} & 0 \\
\rho_{21} & d_2 + \dfrac{nh}{4}
\end{pmatrix},
\]
where the $d_1$ is def\/ined by \eqref{eqnu} and $d_2$ is determined by
\begin{gather}\label{dcond1}
d_1 + d_2 + a_1 + a_2 + a_3 + a_4 +a_5 + a_6 = 0.
\end{gather}
The additional variable, labeled $\rho_{12}$, is
\begin{gather*}
\rho_{21} =  \bar{u}^2 v +\bar{u} \left(a_3-a_1+\frac{n h}{4}+u v\right)
  + u \left(a_3+a_4+a_6+d_1+\frac{n h}{4}\right)+\left(d_1-d_2\right) w_0,
\end{gather*}
however, given that $\mathcal{A}_{2,n}$ is a lower triangular matrix, and that $\mathcal{A}_{3,n} = I$, we may transform our linear problem, $Y_n(x)$, by multiplication on the left by a simple lower triangular matrix so that~$\mathcal{A}_{2,n}$ is may be taken to be diagonal, hence, $\mathcal{A}_n(x)$ takes the general form
\begin{gather}
 A_n(x) =x^3I+  \!\begin{pmatrix} \! \left(d_1 \!+\! \dfrac{nh}{4}\right)((x-\alpha)(x-y) + z_1) \!\!\!\! & \left(d_1 \!+\! \dfrac{nh}{4}\right)\omega(x-y_n) \\ \!\!\!\!\left(d_2 \!+ \!\dfrac{nh}{4}\right)\dfrac{(\delta x-\epsilon)}{\omega} & \left(d_2\! +\! \dfrac{nh}{4}\right)((x-\beta)(x-y)+z_{2})
\!\end{pmatrix}\! , \!\!\!\label{borodin}
\end{gather}
where the function $\omega$ is related to the gauge freedom. The functions, $\alpha$, $\beta$, $\delta$ and $\epsilon$ are specif\/ied by conditions \eqref{dcond1} and \eqref{detrel}. There is also a relation between $z_1$ and $z_2$, which means that $z_1$ and $z_2$ may be written in terms of a single variable, $z$, chosen later to simplify the evolution equations.

In the interest of being explicit, we give expressions for these functions; we def\/ine the notation
\[
\sum_{k=0}^6 \mu_i x^k = \det A_n(x),
\]
then the functions $\alpha$, $\beta$, $\gamma$ and $\delta$ are given, in terms of the $\mu_i$, as
\begin{subequations}\label{parameters}
\begin{gather}
\alpha =  \frac{h^2 n^2}{16 \left(d_1-d_2\right)}+\frac{4 \left(\left(d_2-d_1\right) \left(y^2-z_2\right)+\mu _3+\mu _4
   y\right)}{\left(d_1-d_2\right) \left(4 d_1+n h\right)}\nonumber\\
\hphantom{\alpha =}{} +\frac{n h \left(d_1+d_2-y\right)}{4 \left(d_1-d_2\right)}-\frac{d_1
   \left(y-d_2\right)+\mu _4-2 y^2+z_1+z_2}{d_1-d_2},\\
\beta =  \frac{h^2 n^2}{16 \left(d_2-d_1\right)}-\frac{4 \left(\left(d_1-d_2\right) \left(y^2-z_1\right)+\mu _3+\mu _4
   y\right)}{\left(d_1-d_2\right) \left(4 d_2+n h\right)}\nonumber \\
\hphantom{\beta =}{} -\frac{n h \left(d_1+d_2-y\right)}{4 \left(d_1-d_2\right)}+\frac{y
   \left(d_2-2 y\right)-d_1 d_2+\mu _4+z_1+z_2}{d_1-d_2},\\
\delta =  \alpha  \beta -\frac{4 \left(\mu _0+\mu _1 y\right)}{\left(d_1-d_2\right) y^2 \left(4 d_1+n h\right)}+\frac{4 \left(\mu _0+\mu
   _1 y\right)}{\left(d_1-d_2\right) y^2 \left(4 d_2+n h\right)}\nonumber \\
\hphantom{\delta =}{}
    -\frac{z_1 z_2}{y^2}+y (\alpha +\beta )+z_1+z_2,\\
\epsilon = \frac{16 \mu _0-\left(4 d_1+n h\right) \left(4 d_2+n h\right) \left(\alpha  y+z_1\right) \left(\beta  y+z_2\right)}{y \left(4
   d_1+n h\right) \left(4 d_2+n h\right)}.
\end{gather}
\end{subequations}
The relationship between $z_1$ and $z_2$ is expressed as the determinant
\begin{gather*}
\det A(y) =  \left(y^3 + z_1\left(d_1 + \dfrac{n h}{4}\right)\right)\left(y^3 + z_2\left(d_2 + \dfrac{n h}{4}\right)\right),
\end{gather*}
which we specify by letting the $(1,1)$ and $(2,2)$ entries of $A_n(y_n)$ be
\begin{subequations}\label{zdef}
\begin{gather}
y^3 + z_1 \left(d_1+ \dfrac{nh}{4}\right) = \dfrac{\left(y-a_3\right) \left(y-a_4\right) \left(y-a_5\right) \left(y-a_6\right)}{z+y},\\
y^3 + z_2 \left(d_2+ \dfrac{nh}{4}\right)=  (z+y) \left(y-a_1+\dfrac{nh}{4}\right) \left(y-a_2+\dfrac{nh}{4}\right).
\end{gather}
\end{subequations}
This specif\/ies $z$ in a manner that simplif\/ies the resulting evolution equations. The equations \eqref{borodin}, \eqref{parameters} and \eqref{zdef} specify a parameterization of the Lax matrix described in the work of Arinkin and Borodin \cite{ArinkinBorodin} (written as $\mathcal{A}(z)$ in \cite{ArinkinBorodin}).

We are now in a better position to explicitly relate the system def\/ined by \eqref{systemuv} with \eqref{dPE6}. The variable $y$ in \eqref{dPE6} is
\begin{gather}
\label{yval}y =
v \bar{u}+a_3+u v
 -\frac{v \left(v\bar{u}+u v+a_3-a_4\right) \left(u \bar{v}-v \bar{u}-a_3+a_5\right)}{v \left(2 u \bar{v}- a_2-
   a_3+ a_4+ a_5+\frac{n h}{4}\right)+\bar{v} \left(u \bar{v}- a_2+ a_5+\frac{n h}{4}\right)-v^2 \bar{u}}  ,\!\!\!\!
\end{gather}
and the variable $z$ in \eqref{dPE6} is most succinctly expressed in terms of $y$ as
\begin{gather}
\label{zval}
z+y = \frac{ v \left(y-a_4\right) \left(y-a_5\right) \left(y-a_6\right)}{\left(y- a_1+\frac{n h}{4}\right) \left(v \left( u \bar{v}- a_2+\frac{h
   n}{4}+y\right)+\bar{v} \left(u \bar{v}-a_2+ a_5+\frac{n h}{4}\right)\right)}.
\end{gather}
With these variables, def\/ined\footnote{Equivalent formulations in terms of $\nu$ are easy to derive, but they were not succinct enough for presentation.} in terms of $u$ and $v$, we can verify that the $y$ and $z$ satisfy
\begin{subequations}\label{dPE6bar}
\begin{gather}
 (\bar{\bar{\bar{\bar{z}}}} + y)(y + z)= \dfrac{(y-a_3)(y-a_4)(y-a_5)(y-a_6)}{\left(y-a_1+\frac{nh}{4}\right)\left(y-a_2+\frac{nh}{4}\right)},\\
 (\bar{\bar{\bar{\bar{y}}}} + \bar{\bar{\bar{\bar{z}}}})(y + \bar{\bar{\bar{\bar{z}}}})\nonumber\\
 \qquad {} =  \dfrac{(\bar{\bar{\bar{\bar{z}}}}+a_3)(\bar{\bar{\bar{\bar{z}}}}+a_4)(\bar{\bar{\bar{\bar{z}}}}+a_5)(\bar{\bar{\bar{\bar{z}}}}+a_6)}{\left(\bar{\bar{\bar{\bar{z}}}} -(d_1+a_1+a_2) + \frac{nh}{4}\right)\left(\bar{\bar{\bar{\bar{z}}}} +(d_1+a_3+a_4+a_5+a_6+h) + \frac{nh}{4}\right)} ,
\end{gather}
\end{subequations}
which may be identif\/ied with \eqref{dPE6} when $\delta = h$, $\tilde{f} = \bar{\bar{\bar{\bar{f}}}}$, $t = nh/4$ and
\[
a_7 = - d_1 - a_1 - a_2 , \qquad a_8 = d_1 + a_3 + a_4 + a_5 + a_6 + h.
\]
The constraint, \eqref{constraint}, is trivially satisf\/ied by this choice. At this point, we note that it is somewhat remarkable that this choice, coupled with the def\/inition of the evolution of~$u$ and~$v$, given by~\eqref{systemuv}, are suf\/f\/icient to check that~\eqref{dPE6} is satisf\/ied. To actually derive~\eqref{dPE6}, we require a little more work.

This is a system of linear dif\/ference equations admitting two fundamental solutions, $Y_{\infty}$~and $Y_{-\infty}$, which are analytic throughout the complex plane, except for possible poles to the left and right of the~$a_i$ respectively. The solutions are asymptotically represented, \'a la Birkhof\/f \cite{Birkhoff}, by expansions of the form
\begin{gather*}%\label{series}
Y_{\pm \infty}(x) = x^{3x/h} e^{-3x/h} \left( I + \dfrac{Y_1}{x} + \dfrac{Y_2}{x^2} + \cdots \right) \mathrm{diag} \left(x^{\frac{n-6}{4} + \frac{d_1}{h}},x^{\frac{n-6}{4} + \frac{d_2}{h}}\right).
\end{gather*}
A more current exposition detailing the existence of a more general class of solutions were obtained by \cite{Praagman:Solutions}. The connection matrix is def\/ined by
\[
P(x) = (Y_{\infty}(x))^{-1} Y_{-\infty}(x),
\]
which is $h$-periodic in $x$ (i.e., $P(x+h) = P(x)$). Borodin specif\/ied a canonical group of transformations, specif\/ied by a general form
\begin{gather*}
\tilde{Y}_{n}(x) = R_n(x) Y_n(x),
\end{gather*}
that preserves this matrix \cite{Borodin:connection} which he used with Arinkin to formulate the original Lax pair in~\cite{ArinkinBorodin}.

Since the shift $n \to n+1$ has an undesired non-trivial ef\/fect on the parameters, we need to consider a transformation that sends $n \to n+4$, which we label $B'_n(x)$. A Lax matrix that represents the shift $(l,m) \to (l,m+2)$ may be given by
\begin{gather}\label{prodBp}
B'_n(x) \cong M_{l,m+1}M_{l,m},
\end{gather}
which, is of the general form
\begin{gather*}
B'_n(x) =
\begin{pmatrix}
 x+r_1 & r_2 \vspace{1mm}\\
 -\dfrac{\left(\frac{n h}{4}-a_1-r_1\right) \left(\frac{n h}{4}-a_2-r_1\right)}{r_2} & \dfrac{n h}{2}+x-a_1-a_2-r_1
\end{pmatrix}.
\end{gather*}
This means that our new auxiliary linear equation is
\begin{gather*}%\label{YauxB}
Y_{n+4}(x) = B'_n(x)Y_n(x),
\end{gather*}
where the new compatibility condition is
\begin{gather}\label{compaux}
B'_n(x+h)A_n(x) - A_{n+4}(x) B'_n(x) \equiv 0.
\end{gather}
One may derive the entries, $r_1$ and $r_2$, by performing the required change of variables, from $w_0$ and $w_1$ to functions in $y$ and $z$ to \eqref{prodBp}, however, it is much more convenient to use the \eqref{compaux} directly, to give
%\begin{subequations}
\begin{gather*}
r_1 =  \frac{1}{h}\left(d_1+\frac{n h}{4}\right) (\tilde{\alpha}-\alpha+\tilde{y}-y+1)+\tilde{\alpha} +\tilde{y},\\
r_2 =  \frac{\left(d_1 + \frac{n h}{4}\right) w - \left(d_1 + \frac{n h}{4} + h\right)\tilde{w}}{d_1-d_2+h}.
\end{gather*}
%\end{subequations}
Using the def\/inition of $y$ and $z$, the evolution, and the expressions for $y$ and $z$ in terms of $w_0$ and $w_1$ recovers the expression for $B_n'(x)$ in terms of $w_0$ and $w_1$. Using these values for~$r_1$ and~$r_2$ in \eqref{compaux} gives us \eqref{dPE6}, meaning that the expressions, \eqref{yval} and~\eqref{zval}, in terms of~$u$ and~$v$ solve~\eqref{dPE6bar}.

\section{Symmetries}\label{sec:sym}

The B\"acklund transformations of \eqref{dPE6} form a realization of an af\/f\/ine Weyl group of type $E_6^{(1)}$~\cite{Sakai:rational}. It is interesting and possible to explore how the Schlesinger transformations, such as those explored by Borodin \cite{Borodin:connection}, are tied to the concept of consistency around a cube \cite{Nijhoff:CAC}. That is, we seek to relate the B\"acklund transformations for lattice equation, def\/ined by the multilinear function, \eqref{H1}, to the discrete isomonodromic (or connection preserving) deformations of Borodin \cite{Borodin:connection}.

Naturally, the connection preserving deformations form sub-groups of the full af\/f\/ine Weyl groups of symmetries. The $q$-analogues of connection preserving deformations appeared in the work of Jimbo and Sakai \cite{Sakai:qP6}, and there have been but a few steps towards a general system of Schlesinger transformations that describe symmetries for the discrete Painlev\'e equations \cite{Ormerodlattice, Sakai:Schlesinger}.

The f\/irst set of symmetries we wish to describe is permutations on the set of parameters $\{a_1,a_2\}$ and the permutations on the set $\{a_3,a_4,a_5,a_6\}$. Let us f\/irst describe the symmetry $s_{34}$, which permutes $a_3$ and $a_4$, which we will generalize to provide the symmetry for the remaining symmetry group on the set of parameters. It is important to also note that these symmetries commute with the fourth power of the generating shift, not the generating shift itself.

\begin{figure}[t]\centering
\includegraphics{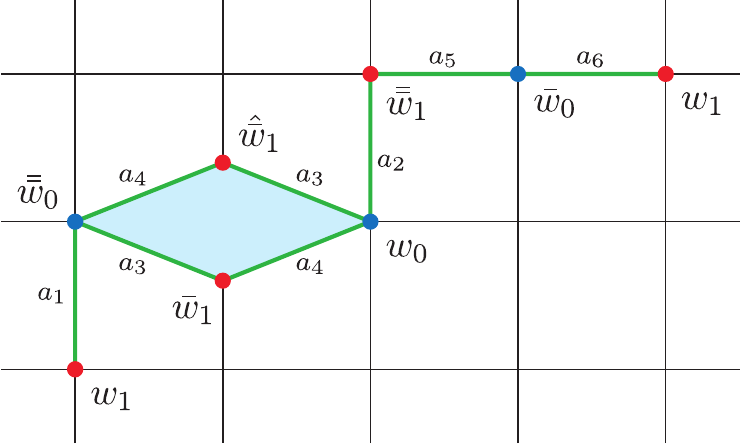}
\caption{A pictorial representation of the symmetry $s_{34}$ which permutes the roles of $a_3$ and $a_4$.}\label{fig:s34}
\end{figure}

Let us describe the symmetry $s_{34}$ as a map that sends the lattice variables, $w_i$ and their iterates, to $\hat{w}_i$. The path described in \ref{labelling} assumes a certain labeling where, left to right, the parameters cycle through $a_3$ to $a_6$. The action of $s_{34}$ on the $w_i$ has a trivial ef\/fect on almost all the lattice variables except for $\bar{w}_1$, which is related to $\hat{\bar{w}}_1$ by a quad centered at $\bar{w}_1$ with variables $a_3$ and $a_4$ on the edges, as in Fig.~\ref{fig:s34}.

That is to say that the symmetry is described by the equation
\[
Q(\bar{\bar{w}}_0, \bar{w}_1, \hat{\bar{w}}_1, w_0, a_4,a_3) = 0,
\]
which means that the $s_{34}$ has the ef\/fect
\[
s_{34} : \  \bar{w}_1 \to \hat{\bar{w}}_1 = \bar{w}_1 + \dfrac{a_3 -a_4}{w_0 - \bar{\bar{w}}_0}, \qquad s_{34} : \ a_3 \to a_4, \qquad s_{34} : \ a_4 \to a_3.
\]
This may be easily lifted to a symmetry of the $u$ and $v$ variables, where we note that the~$u$ and~$\bar{u}$ are unchanged, and the $v$ variables become{\samepage
\begin{gather*}
s_{34} : \ v \to \hat{v} = v + \dfrac{a_3-a_4}{u + \bar{u}}, \qquad s_{34} : \ \bar{v} \to \hat{\bar{v}} = \bar{v} - \dfrac{a_3-a_4}{u + \bar{u}}.
\end{gather*}
This allows us to easily show that $\hat{y} = y$ and $\hat{z} = z$.}

Similarly, to obtain the transformation, $s_{56}$, on the lattice variables, we place a quad centered at $\bar{w}_0$, which will permute $a_5$ and $a_6$ in the same way. That is to say that $s_{56}$ has a trivial ef\/fect on all the lattice variables except for $\bar{w}_0$, which is related to the transformed variable, $\hat{\bar{w}}_0$ via the relation
\[
Q(\bar{\bar{w}}_1, \hat{\bar{w}}_0,\bar{w}_0, w_1, a_5,a_6) = 0,
\]
which means $s_{56}$ is specif\/ied by
\[
s_{56} : \ \bar{w}_0 \to \hat{\bar{w}}_0 = \bar{w}_0 + \dfrac{a_5 -a_6}{w_1 - \bar{\bar{w}}_1}, \qquad s_{56} : \ a_5 \to a_6, \qquad s_{56} : \ a_6 \to a_5.
\]
We lift this to the variables $u$ and $v$ to give a trivial ef\/fect on $v$ and an action on $u$ and $\bar{u}$ described by
\[
s_{56} : \ u \to \hat{u} = u + \dfrac{a_5-a_6}{v + \bar{v}}, \qquad s_{56} : \ \bar{u} \to \hat{\bar{u}} = \bar{u} - \dfrac{a_5-a_6}{v + \bar{v}},
\]
while, once again, this change has a trivial ef\/fect on $y$ and $z$.

\begin{figure}[t]\centering
\includegraphics{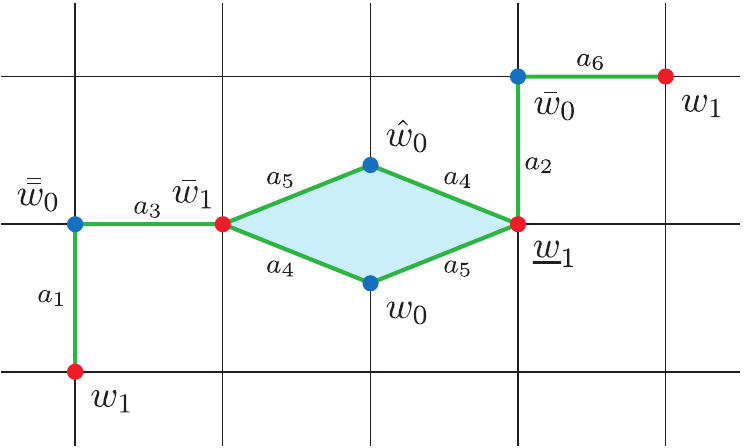}
\caption{A pictorial representation of the symmetry $s_{45}$ which permutes the roles of $a_4$ and $a_5$.}\label{fig:s45}
\end{figure}

To extrapolate this principle to obtain $s_{12}$ and $s_{45}$, we remark that the while the path chosen above gives us natural places to insert quads that permute the parameters, the path chosen is still arbitrary. That is to say that because $Q$ is multilinear, specifying the six initial conditions on any path is in one-to-one correspondence with the specif\/ication of variables along any other path. In this sense, while the standard staircase of \cite{dKdVreds, vdKRQ:Staircase} is useful in describing the particular form of the evolution equations, \eqref{systemw}, it is not actually that important in building a correspondence between~\eqref{systemw} or \eqref{systemuv} with~\eqref{dPE6}. With this in mind, let us consider a small deviation of our original path that passes through $\underline{w}_1$ instead of $\bar{\bar{w}}_1$, as depicted in Fig.~\ref{fig:s45}.

The quad centered at $w_0$ in Fig.~\ref{fig:s45} has a trivial ef\/fect on each of the variables except for~$w_0$. Knowing $\hat{w}_0$ and $\hat{\underline{w}}_1 = \underline{w}_1$ is suf\/f\/icient to obtain $\hat{\bar{\bar{w}}}_1$, i.e., one can recover the transformation of the variables along our original path by using the quad equation. This gives us that the symmetry that permutes the~$a_4$ and~$a_5$ is trivial for all variables except for $w_0$ and $\bar{\bar{w}}_1$, which is related to the transformed variables by the two quad equations
\begin{gather*}
Q(\bar{w}_1, \hat{w}_0, w_0, \underline{w}_1,a_5,a_4)  = 0,\qquad
Q\left(\hat{w}_0, \underline{w}_1, \hat{\bar{\bar{w}}}_1, \bar{w}_0,a_2-\frac{nh}{4},a_4\right)  = 0,
\end{gather*}
where the $nh/4$ contribution comes from the dif\/ference between an $p_l$ and a $q_m$ and $\underline{w}_1$ is known from knowledge of the original initial conditions, which gives us
\begin{gather*}
s_{45}  : \ w_0 \to \hat{w}_0 = w_0 +  \frac{\left(a_4-a_5\right) \left(w_0-\bar{w}_0\right)}{a_5- a_2+\frac{nh}{4}+ \left(w_0-\bar{w}_0\right)
   \left(\bar{w}_1-\bar{\bar{w}}_1\right)},\\
s_{45}  : \ \bar{w}_1 \to \hat{\bar{w}}_1 = \bar{w}_1 + \dfrac{(w_0 - \bar{w}_0)(\bar{w}_1 - \bar{\bar{w}}_1)}{\bar{w}_0-\hat{w}_0}, \qquad s_{45} : \ a_4 \to a_5,\qquad s_{45} : \ a_5 \to a_4.
\end{gather*}
The transformation f\/ixes $\bar{u}$ and $v$, and changes $u$ and $\bar{v}$ in the following manner,
\[
s_{45}: \ u \to \hat{u} = u + \dfrac{u(a_4-a_5)}{\frac{n h}{4} + u \bar{v} - a_2 +a_5}, \qquad s_{45} : \ \bar{v} \to \hat{\bar{v}} = \bar{v} + \dfrac{\bar{v}(a_4-a_5)}{\frac{n h}{4} + u \bar{v} - a_2 + a_4},
\]
which also has a trivial ef\/fect on $y$ and $z$.

To obtain the transformation that permutes the $a_1$ and $a_2$ variables, the simplest choice of path passes through $\tilde{w}_1$ (i.e., $w_1(n+4)$) and $\bar{\bar{\bar{w}}}_3$. By placing a quad centered at $\bar{\bar{w}}_0$, the transformation f\/ixes all the variables, except for $\bar{\bar{w}}_0$ on this path, meaning the transformation, changes $w_0$, $\bar{w}_1$ and $\bar{\bar{w}}_1$ in accordance with
\begin{gather*}
Q\left(w_1, \bar{\bar{w}}_0,\hat{\bar{\bar{w}}}_0, \tilde{w}_1, a_2,a_1\right) = 0,\qquad
Q\left(\hat{\bar{\bar{w}}}_0, \hat{\bar{w}}_1, \tilde{w}_1, \bar{\bar{\bar{w}}}_0, a_1-\frac{nh}{4},a_3\right)=0,\\
Q\left(\hat{\bar{w}}_1, \bar{\bar{\bar{w}}}_0,\hat{w}_1, \hat{w}_0, a_1-\frac{nh}{4},a_4\right)=0,
\end{gather*}
which we will write implicitly as
\begin{alignat*}{3}
& s_{12}: \ \bar{\bar{w}}_0 \to \hat{\bar{\bar{w}}}_0 = \bar{\bar{w}}_0 + \dfrac{a_2-a_1}{\tilde{w}_1 - w_1}, \qquad && s_{12} : \ \bar{w}_1 \to \hat{\bar{w}}_1 = \tilde{w}_1 + \dfrac{\frac{n h}{4}-a_1+a_3}{\bar{\bar{\bar{w}}}_0 - \hat{\bar{\bar{w}}}_0}, &\\
& s_{12}: \ w_0 \to \hat{w}_0 = \bar{\bar{\bar{w}}}_0 + \dfrac{\frac{n h}{4}-a_1+a_4}{\bar{\bar{w}}_1 - \hat{\bar{w}}_1}, \qquad && s_{12} : \ a_1 \to a_2, \qquad s_{12} : \ a_2 \to a_1,&&
\end{alignat*}
which again, may be expressed in terms of transformations in $u$ and $v$ as
\begin{alignat*}{3}
& s_{12} : \ u \to \hat{u} = u_n + \dfrac{\frac{n h}{4} - a_2+a_4}{\bar{v}} - \dfrac{\frac{nh}{4}-a_1+a_4}{\hat{\bar{v}}},  \qquad && s_{12} : \ v \to \hat{v} = v+ \lambda,& \\
& s_{12} : \ \bar{u} \to \hat{\bar{u}} = \bar{u} + \dfrac{\frac{n h}{4} - a_1 + a_3}{v} - \dfrac{\frac{n h}{4} - a_2+a_3}{\hat{v}}, \qquad && s_{12} : \ v \to \hat{\bar{v}} = \bar{v}- \lambda, &
\end{alignat*}
where{\samepage
\[
\lambda = \frac{\left(a_1-a_2\right) v \bar{v}}{v \left(\left(\bar{u}+u\right) \bar{v}- a_2+ a_4+\frac{n h}{4}\right)+\bar{v} \left(a_3- a_1+\frac{n h}{4}\right)}.
\]
This transformation has a trivial ef\/fect on $d_1$, $d_2$, $y$ and $z$.}

Let us now describe two transformations, $T_{a_1}$ and $T_{a_3}$, which, in combination with the symmetries described above, will be suf\/f\/icient to describe a complete set of connection preserving deformations described by Borodin \cite{Borodin:connection}. These transformations will necessarily act non-trivially on the variable $y$ and $z$ and may be used to obtain inf\/inite families of rational and Riccati solutions from the seed solutions in Section~\ref{sec:sols}.

The f\/irst transformation, $T_{a_1}$, may be considered to be the image of the shift in the positive $m$ direction. This may also be seen to be induced by a lift of the $M_{l,m}$ matrix, \eqref{LaXM}, to the level of the reduction. This also represents the discrete analogue of a fundamental Schlesinger transformation in the work of Jimbo and Miwa \cite{Jimbo:Monodromy2} since \eqref{LaXM} induces the transformation of the linear system
\begin{gather}\label{Ra1}
\hat{Y}_n(x) = R_{n,a_1}(x) Y_n(x),
\end{gather}
where $R_{n,a_1}(x)$ is linear in $x$ and
\[
\det R_{n,a_1}(x) = x-a_1 - \frac{nh}{4}.
\]
This transformation may be described on the level of the lattice variables by
\begin{alignat*}{3}
&  T_{a_1} : \ w_1 \to \hat{w}_1 = \bar{\bar{w}}_0, \qquad && T_{a_1} : \ w_0 \to \hat{w}_0 = \bar{\bar{w}}_1, & \\
&Q\left(\bar{w}_1, w_0, \hat{\bar{w}}_1, \bar{\bar{w}}_1, a_2 - \frac{n h}{4},a_4\right) = 0, \qquad && Q\left(\bar{w}_0, w_1, \hat{\bar{w}}_0, \bar{\bar{w}}_1, a_1+h - \frac{n h}{4},a_5\right) = 0,& \\
&Q\left(\bar{\bar{w}}_0, \bar{w}_1, \hat{\bar{\bar{w}}}_0, \hat{\bar{w}}_1, a_2 - \frac{n h}{4},a_3\right) = 0,\qquad && Q\left(\bar{\bar{w}}_1, \bar{w}_0, \hat{\bar{\bar{w}}}_1, \hat{\bar{w}}_0, a_1+h - \frac{n h}{4},a_6\right) = 0,&
\end{alignat*}
where the parameters, $a_1$ and $a_2$ change via the rule
\[
T_{a_1} : \ a_1 \to a_2, \qquad T_{a_1} : \ a_2 \to a_1 - h.
\]
When lifted to the level of the $u$ and $v$ variables, the transformation is described by
\begin{gather*}
T_{a_1}  : \ u \to \hat{u} = -v - \bar{v} - \dfrac{\frac{h(n+4)}{4} -a_1 + a_6}{\bar{u}}, \\
T_{a_1}  : \ v \to \hat{v} = -u - \bar{u} - \dfrac{\frac{n h}{4} - a_2+ a_4}{\bar{v}},\\
T_{a_1}  : \ \bar{u} \to \hat{\bar{u}} = v + \dfrac{\frac{h(n+4)}{4} - a_1 + a_6}{\bar{u}} - \dfrac{\frac{nh}{4}-a_2+a_3}{\hat{\bar{v}}},\\
T_{a_1}  : \ \bar{v} \to \hat{\bar{v}} = u - \dfrac{\frac{h(n+4)}{4} - a_1 + a_5}{\hat{u}} + \dfrac{\frac{nh}{4}-a_2+a_4}{\bar{v}},
\end{gather*}
which is equivalent to a double shift in $n$. The f\/irst implication of this is that our invariant, $d_1$~and $d_2$ change, in accordance with the rule
\[
T_{a_1} : \ d_1 \to d_2, \qquad T_{a_1} : \ d_2 \to d_1 + h.
\]
To calculate the ef\/fect of $T_{a_1}$ on $y$ and $z$, we use the compatibility between \eqref{Ra1} and \eqref{Ydiff}, where $A_n(x)$ is def\/ined by \eqref{borodin}. The ef\/fects of $T_{a_1}$ on $y$ and $z$ are given by
\begin{gather*}
T_{a_1}  : \ y  \to \hat{y} =y + \dfrac{1}{d_1-d_2} \left(y^2 + \frac{\delta  \left(d_1+\frac{n h}{4}\right) \left(d_2+\frac{n h}{4}\right)}{\left(d_1+\frac{n h}{4}\right) \left(\alpha - a_1+\frac{n h}{4}\right)-\frac{\left(\frac{n h}{4}- a_1\right)^3- z_1 \left(d_1+\frac{n h}{4}\right)}{y- a_1+\frac{n h}{4}}} \right. \\
 \left.\qquad{} -\frac{z_1 \left( d_1+\frac{n h}{4}\right)+ y^3}{y- a_1+\frac{n h}{4}}+a_1 \left(d_2+a_1-\frac{n h}{4}+y\right)-d_2 \left(\beta +\frac{n h}{4}\right)-\frac{(\beta+y)  n h}{4} \right),\\
T_{a_1}  : \ z \to \hat{z},
\end{gather*}
where the relation between $z$ and $\hat{z}$ is given implicitly by the relation
\begin{gather*}
\hat{z}_1 =  \dfrac{y-\hat{y}}{d_2+\frac{n h}{4}} \left(d_1 \left( \alpha - a_1+\frac{n h}{4}\right)- \frac{a_1 n h}{4}+ \left(\beta -\hat{y}\right) \left(d_2+\frac{n h}{4}\right)+\frac{h^2 n^2}{16}+\frac{\alpha  n h}{4} \right.\\
 \left. \hphantom{\hat{z}_1 =}{} -\frac{\left(\frac{n h}{4}-a_1\right){}^3- z_1 \left(d_1+\frac{n h}{4}\right)}{ \left(- a_1+\frac{n h}{4}+
   y\right)}+\frac{z_2 \left( d_2+\frac{n h}{4}\right)}{ \left(y-\hat{y}\right)} \right).
\end{gather*}

Similarly, if we were to lift the $L_{l,m}$ to the reduced linear problem, we obtain a linear system
\begin{gather}\label{Ra3}
\hat{Y}_n(x) = R_{n,a_3}(x)Y_n(x),
\end{gather}
where $R_{n,a_3}(x)$ is also linear in $x$ and
\[
\det R_{n,a_3}(x) = x-a_3.
\]
This is a fundamental Schlesinger transformation that induces a transformation, which we label~$T_{a_3}$, that has the ef\/fect of f\/ixing $a_1$ and $a_2$, permuting the other variables, $a_3$ to $a_6$, as follows
\begin{gather*}
T_{a_3} : \ a_3 \to a_4, \qquad T_{a_3} : \ a_4 \to a_5,\qquad T_{a_3} : \ a_5 \to a_6, \qquad T_{a_3} : \ a_6 \to a_3 -h.
\end{gather*}
The ef\/fect on the lattice variable is given in terms of the multilinear function, $Q$, as
\begin{alignat*}{3}
& T_{a_3} : \ \bar{w}_0 \to \hat{\bar{w}}_0 = w_1, \qquad && T_{a_3} : \ \bar{w}_1 \to \hat{\bar{w}}_1 = w_0, & \\
& T_{a_3} : \ \bar{\bar{w}}_0 \to \hat{\bar{\bar{w}}}_0 = \bar{w}_1, \qquad && T_{a_3} : \bar{\bar{w}}_0 \to \hat{\bar{\bar{w}}}_1 = \bar{w}_0,& \\
&Q\left(w_1,\hat{w}_1,\bar{\bar{w}}_0,\bar{w}_1,a_3,a_1 - \frac{n h}{4}\right) = 0, \qquad && Q\left(w_0,\hat{w}_0,\bar{\bar{w}}_1,\bar{w}_0,a_5,a_2 - \frac{n h}{4}\right) = 0.&
\end{alignat*}
The ef\/fect of $T_{a_3}$ on $u$ and $v$ is simply given by the inverse of \eqref{systemuv}. In the same way as for $T_{a_1}$, we use \eqref{Ra3}, \eqref{Ydiff} and \eqref{borodin} to f\/ind that
\[
T_{a_3} : \ d_1 \to d_2, \qquad T_{a_3} : \ d_2 \to d_1 + h,
\]
and the ef\/fect on $y$ and $z$ is given by
\begin{gather*}
T_{a_3} : \ y  \to\hat{y} =  y+ \dfrac{1}{d_1-d_2}\left (\frac{z_1 \left(d_1+\frac{n h}{4}\right)+ y^3}{\left(a_3-y\right)}+y^2-\beta d_2+a_3 \left( d_2+\frac{n h}{4}+y\right)
 \right.\\
  \left.\qquad{} -\frac{\beta  n h}{4}+a_3^2 +\frac{\delta  \left(y-a_3\right) \left(d_1+\frac{n h}{4}\right) \left(d_2+\frac{n h}{4}\right)}{\left(
   \left(d_1+\frac{n h}{4}\right)\left(y \left(\alpha -a_3\right)-\alpha  a_3 +z_1\right)+a_3^2 \left(a_3+ d_1+\frac{n h}{4}\right)\right)} \right),\\
T_{a_3} : \ z  \to \hat{z},
\end{gather*}
where we give the relation implicitly by stating that $\hat{z}_1$ is specif\/ied by
\begin{gather*}
\hat{z}_1 =  \dfrac{y-\hat{y}}{d_2+\frac{n h}{4}} \left(\frac{z_1 \left(d_1+\frac{n h}{4}\right)+ y^3}{\left(y-a_3\right) \left(d_2+\frac{n h}{4}\right)}-\frac{a_3 \left(d_1+\frac{n h}{4}+y\right)}{d_2+\frac{n h}{4}}-\frac{a_3^2}{d_2+\frac{n h}{4}}+\beta \right. \\
  \left.\hphantom{\hat{z}_1 =}{} +\frac{\alpha  \left(d_1+\frac{n h}{4}\right)- y^2}{d_2+\frac{n h}{4}}+\frac{z_2}{y-\hat{y}}-\hat{y}\right).
\end{gather*}

What is important in these calculations is that the general symmetry structure of reductions of quad-equations apparent. In fact, the way in which the general symmetry structure is described by quad-equations is much more natural as a~higher-dimensional reduction than a~two-dimensional reduction. In this case, this is more naturally a 6-dimensional reduction. This work, in combination with \cite{OvdKQ:reductions}, is suf\/f\/icient to give a symmetry structure to reductions that are equivalent to all the systems below those with~$A_2^{(1)}$ surfaces in the Sakai classif\/ication~\cite{Sakai:rational}.

It should be noted that the above constitutes a set of symmetries that have been derived by an application of $Q$ as a symmetry. There are other equations that are consistent with $Q$, which can be found in the work of Boll~\cite{boll2011classification}, however, none of the tested equations that are consistent with~\eqref{H1} seemed to produce a non-trivial transformation of the Painlev\'e variables, $y$~and~$z$. Perhaps some further investigation is required in this direction.

\section{Special solutions}\label{sec:sols}

Lastly, this brings us to the method we wish to use to solve~\eqref{dPE6}. There are numerous studies that show that the Painlev\'e equations admit solutions expressible in terms of hypergeometric and conf\/luent hypergeometric functions (see~\cite{Noumi:survey} for a review). Many of the simplest discrete Riccati solutions were, including those for~\eqref{dPE6}, were presented in \cite{ramani2001special}. A more geometric approach was taken for the $q$-Painlev\'e equations by Kajiwara et al.~\cite{qPhypergeometric}. The work of Nicholas Witte has shown that is possible to f\/ind many hypergeometric solutions, including those for~\eqref{dPE6}, in terms of the moments of a certain semi-classically deformed orthogonal polynomial ensembles~\cite{Witte:Correspondence}, which is closely related to the Pad\'e approximation method of Yamada et al.~\cite{yamada2009pade}.

For an equation such as \eqref{dPE6}, it is not immediately obvious how to even obtain very basic rational solutions. Our work on the reduction gives us a number of dif\/ferent equivalent forms of the equation that we may solve explicitly. If we take \eqref{systemuv} in the special case in which the parameters of the lattice equation, $p_l$ and $q_m$, are linear
\begin{gather*}
p_l = a + \dfrac{hl}{4}, \qquad q_m = b + \dfrac{hm}{2},
\end{gather*}
corresponding to the choice of parameters
\begin{gather*}
a_1 = -b, \quad a_2 = -b - \frac{h}{2}, \quad a_3 = -a, \quad a_4 = -a - \frac{h}{4}, \quad a_5 = -a - \frac{h}{2}, \quad a_6 = -a - \frac{3h}{4},
\end{gather*}
the evolution equations simplify to
\begin{gather*}
\bar{v}(u + \bar{u} + \bar{\bar{u}})  = a-b- \dfrac{(n+1)h}{4},\qquad
\bar{u}(v + \bar{v} + \bar{\bar{v}})  = a-b- \dfrac{(n+1)h}{4}.
\end{gather*}
If $u = v$, this becomes a version of d-$\mathrm{P}_{\rm I}$ that has no rational solutions. This equation does possess a relatively simple one parameter family of rational solutions, in the case that~$u$ is constant and~$v$ is linear, explicitly we have
\begin{gather*}
u  = \dfrac{1}{3\lambda},\qquad
v  = \lambda\left(a-b + \frac{n h}{4}\right),
\end{gather*}
or equivalently, this choice prescribes a two parameter family of solutions to the system in~$w_0$ and~$w_1$, given by
\begin{gather*}
w_0  = \dfrac{n}{3\lambda} +\theta_1,\qquad
w_1  = \lambda n\left(a-b + \frac{h}{8}\right) + \dfrac{\lambda n^2h}{8} + \theta_2,
\end{gather*}
where $\theta_1$ and $\theta_2$ are arbitrary constants. Substituting these solutions into $y$, $z$, $d_1$ and $d_2$ gives
\begin{gather*}
d_1  = 2a+b + \dfrac{5h}{6} , \qquad d_1 = 2a+b + \dfrac{7h}{6},\\
y =  \frac{1}{24} \bigg(\frac{7 h^2 (2 n h-h-8 a+8 b)}{64 a^2-16 a (8 b+h (2 n-1))+64 b^2+16 b h (2 n-1)+h^2 (4 (n-1) n-1)}  \\
\hphantom{y =}{}  -8 a-16 b-4 h
   n-7 h\bigg),\\
z =  \frac{1}{24} \left(h \left(\frac{14 h}{-8 a+8 b+2 n h-5 h}-4 n+19\right)+40 a-16 b\right).
\end{gather*}
Alternatively, by swapping the roles of $u$ and $v$ we obtain the solution
\begin{gather*}
d_1  = 2a+b + \dfrac{2h}{3}, \qquad d_2 = d_2 = 2a+b + \dfrac{4h}{3},   \\
y  = \frac{1}{24} \left(h \left(-\frac{5 h}{8 a-8 b-2 n h+h}-4 n-7\right)-8 a-16 b\right),\\ %\label{ratseedy}
z  = \frac{1}{24} (40 a-16 b+h (19-4 n)). %\label{ratseedz}
\end{gather*}

To consider the hypergeometric solutions, let us f\/irst review the work of Ramani et al.~\cite{ramani2001special} who present solutions of the form
%\begin{subequations}
\begin{gather*}
%\label{ty}
\tilde{y}  = \dfrac{\sigma_1 \tilde{z} + \sigma_2}{\sigma_3 \tilde{z} + \sigma_4},\qquad
%\label{tz}
\tilde{z}  = \dfrac{\tau_1 y + \tau_2}{\tau_3 z + \tau_4},
\end{gather*}
where the $\sigma_i$'s and $\tau_i$'s are functions on $n$ alone. There are a multitude of values the $\sigma$ and $\tau$ variables may take in order for the evolution to coincide with \eqref{dPE6}. We choose just one, which yields
\begin{gather*}
\tilde{y} =\frac{\tilde{z} \left(a_2-\frac{1}{4} h (n+4)\right)+a_3 a_4}{\tilde{z}-a_2+a_3+a_4+\frac{h
   n}{4}+h},\qquad
\tilde{z} =\frac{-y \left( a_6+\frac{n h}{4}\right)+ a_1 y+ a_5 \left(a_6-y\right)}{y- a_1+\frac{n h}{4}},
\end{gather*}
%\end{subequations}
where we have one of two alternative constraints, either
\[
d_1 +a_1 + a_3 + a_4 +h =0\qquad \textrm{or} \qquad d_1 +a_2 + a_3 + a_4 +h =0.
\]
Combining these two equations gives us a single second order dif\/ference equation which also admits the hypergeometric dif\/ferential equation in a continuum limit \cite{ramani2001special}. These solutions may be identif\/ied as hypergeometric functions by relating these recurrence relations to contiguous relations for hypergeometric functions of type ${}_3F_2(1)$, as done by Kajiwara~\cite{kajiwara:hypergeometricEs}.

We can now consider the possibility of the evolution of $\nu$ coinciding with a fractional linear transformation of the form
\begin{gather}\label{fracline}
\bar{\nu} = \dfrac{c_{11} \nu + c_{12}}{c_{21} \nu + c_{22}},
\end{gather}
which may be inverted so calculate $\underline{\nu}$ in terms of $\nu$. Substituting \eqref{fracline}, and its inverse, into \eqref{eqnu} gives us a number of conditions in the variable $\nu$, which we solve to give the equation
\begin{gather*}%\label{Riccati}
\bar{\nu} = \dfrac{(1+\nu)\left( \dfrac{nh}{4}+d_1 + a_3 + a_4 + a_5\right)}{a_2 - a_4 - \dfrac{n h}{4} - \nu\left( \dfrac{n h}{4}+ d_1 +a_3 + a_4 + a_5 \right)},
\end{gather*}
which coincides with the evolution of \eqref{eqnu} so long as
\[
2d_1 + a_1 + a_2 + a_3 + a_4 + a_5 + a_6 = 0.
\]
In the language of the af\/f\/ine Weyl symmetries, this condition ensures that the parameters correspond to a point on a wall of an af\/f\/ine Weyl chamber.

To obtain the corresponding solution for $y$ and $z$, we f\/irst take the fourth power of this mapping to f\/ind $\tilde{nu}$ is given by
\[
\tilde{\nu} = \dfrac{C_{11} \nu + C_{12}}{C_{21} \nu + C_{22}},
\]
where
\begin{gather*}
C_{11} =   \left(\dfrac{n h}{2}- a_1- a_2+ a_3+ a_4+ a_5- a_6\right) \left(\dfrac{h (n+4)}{2}-a_1-a_2+a_3+a_4-a_5+a_6\right), \\
C_{12} =  - \left(2h-a_1+a_2-a_3-a_4+a_5+a_6\right) \left(\dfrac{h (n+4)}{2}-a_1-a_2+a_3+ a_4- a_5+ a_6\right),\\
C_{21} =  \left(a_1-a_2-a_3-a_4+a_5+a_6\right) \left(\dfrac{n h}{2}- a_1- a_2+ a_3+ a_4+ a_5- a_6\right),\\
C_{22} =  \frac{h^2 (n+4)^2}{4} -n h \left(a_1+a_2-a_5-a_6\right) + 2 \left(a_3 h+a_4 h+a_5 h+a_6 \left(a_3+a_4+h\right)\right. \\
 \left. \hphantom{C_{22} =}{}
 -a_2 \left(a_5+a_6+h\right)-a_1 \left(a_5+a_6+3
   h\right) +a_1^2+a_2^2-a_3^2-a_4^2+a_3 a_5+a_4 a_5\right).
\end{gather*}
The constraint seems characteristically dif\/ferent from the solutions above, which means this is a hypergeometric solution for a dif\/ferent choice of parameters. We do not know how these two solutions are related.

Using the group of symmetries derived in Section~\ref{sec:sym} one is able to generate an inf\/inite number of hypergeometric solutions, however, we do note that the one cannot apply the full set of B\"acklund transformations to the hypergeometric solutions. One characterization of the hypergeometric solutions is that they are solutions that are singular on a translational B\"acklund transformation. That is to say that these solutions are not just on a wall of the af\/f\/ine Weyl chamber, but they are on a barrier which one cannot pass through using B\"acklund transformations.

It would be interesting to consider determinantal structures, either from a bilinear approach~\cite{KajiwaraqP3I} or an orthogonal polynomial approach~\cite{OrmerodForresterWitte}. Determinantal formulations are beyond the scope of this paper.

\section{Conclusion}

We have provided a reduction from one of the most degenerate lattice equations to an equation that sits above the sixth Painlev\'e equation in Sakai's classif\/ication. While these reductions are somewhat complicated, their very existence suggests that the solutions to all the additive Painlev\'e equations, and many higher-dimensional additive equations, may be expressed in terms of the solutions of the lattice potential Korteweg--de Vries equation. A similar statement could be made about multiplicative type equations and the discrete Schwarzian Korteweg--de Vries equation. Furthermore, while we know of a reduction from the Schwarzian Korteweg--de Vries equation to the sixth Painlev\'e equation \cite{SKdVP6I, SKdVP6II}, we believe there may exist a more complicated reduction from the Korteweg--de Vries equation to the sixth Painlev\'e equation that is yet to be found.

There is still a sense in which these reductions may be more naturally approached from a~geometric perspective, where one is interested in mapping between biquadratic curves for the lattice equations and the (moving) biquadratics for the (non-autonomous) reductions. While this work suggests that the symmetries are most naturally viewed in terms of quad-equations on higher-dimensional lattices, Doliwa notes in \cite{Doliwa:noncomGDsys} that it might be more natural to consider reductions from higher-dimensional lattice equations.

There are still also issues regarding the def\/initions and framework of connection preserving deformations. These connection preserving deformations should manifest themselves as automorphisms of the Galois group associated with the system of dif\/ference equations. This could present a natural extension of the connection preserving deformations that could be large enough to describe the full af\/f\/ine Weyl group of B\"acklund transformations, rather than just a restriction to the translational elements of the af\/f\/ine Weyl group.

This work on the symmetries of reduced equations applies a wide class of periodic reductions that have appeared in the literature. It is not entirely clear at this point how, or even whether, this procedure may be applied to the so-called twisted reductions explored by recent work~\cite{hay2013systematic, ormerod2013twisted}. This work suggests that the symmetry group of a $(s_1,s_2)$-reduction is at least a lattice of dimension $s_1 + s_2$, hence, we suspect it would take a $(4,4)$-reduction of Q4 to be able to be identif\/ied with the full parameter version of the elliptic Painlev\'e equation.

\subsection*{Acknowledgements}

This research is supported by Australian Research Council Discovery Grant \#DP110100077.

\pdfbookmark[1]{References}{ref}
\LastPageEnding

\end{document}